\documentclass{iau}
\usepackage{graphicx}

\def\deg{$^\circ$}

\title[IAUS291.~~The early science opportunities of FAST] 
{The Five-hundred-meter Aperture Spherical Radio Telescope Project and its Early Science Opportunities} 

\author[D. Li, R. Nan \& Z. Pan]
{Di Li$^1$$^,$$^2$,
\ Rendong Nan$^1$$^,$$^3$
\and Zhichen Pan$^1$}

\affiliation{$^1$National Astronomical Observatories, Chinese Academy of Science, \\ A20 Datun Road, Chaoyang District, Beijing 100012, China \\ email: {\tt dili@nao.cas.cn} \\[\affilskip]
$^2$Space Science Institute, Boulder, CO 80301, USA \\[\affilskip]
$^3$Key Laboratory for Radio Astronomy, Chinese Academy of Sciences, Nanjing 210008, China\\[\affilskip]}

\pubyear{2012}
\volume{291}  
\jname{\mbox{Neutron Stars and Pulsars: Challenges and Opportunities after 80 years}}
\editors{J. van Leeuwen, ed.}
\begin{document}

\maketitle

\begin{abstract}
The National Astronomical Observatories, Chinese Academy of Science (NAOC), has started building the largest
antenna in the world. Known as FAST, the Five-hundred-meter Aperture Spherical radio Telescope  is a Chinese mega-science project funded by the National Development and Reform Commission (NDRC).
FAST also represents part of Chinese contribution to the international efforts to build the square kilometer array (SKA). Upon its finishing around September of 2016, FAST will be the most sensitive single-dish radio telescope in the low frequency radio bands between 70 MHz and 3 GHz. The design specifications of FAST, its expected capabilities, and its main scientific aspirations  were described in an overview paper by Nan et al.~(2011). In this paper, we briefly review the design and the key science goals of FAST, speculate the likely limitations at the initial stages of FAST operation, and discuss the opportunities for astronomical discoveries in the so-called early science phase.
	
\end{abstract}


\firstsection 
\section{The FAST Project}
FAST is an Arecibo-type antenna with three outstanding aspects. First, it is sited in a deep karst depression, Dawodang in Guizhou province in southwestern China, which allows for a 500 meter spherical aperture
and a zenith angle of 40 degrees. Second, the spherical aberration is
to be corrected by an active primary surface comprised of more than 4000 steerable panels. Third,
a light-weight feed cabin will be driven by six cables and a servomechanism plus a parallel robot
to realize close-loop precision control. Compared with Arecibo, these features facilitate
three main advantages, about twice the effective collecting area, about twice the sky coverage, and a much
lighter focal cabin structure and thus a cleaner optical path.
An original image of the site, the schematics of FAST optics, and a 3D model are shown
in Figure 1.
 \begin{figure}[b]
 \begin{center}
 \includegraphics[width=0.9\textwidth]{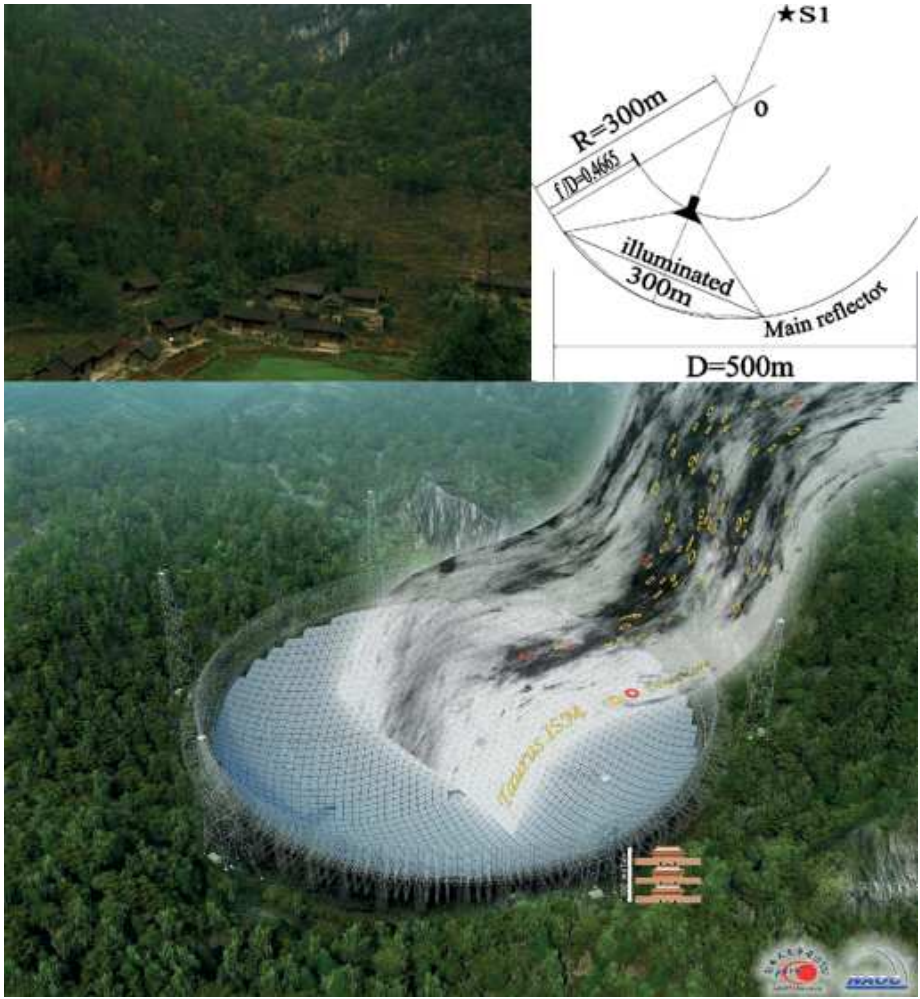}
 \caption{{\em Upper Left:} Partial view of the original site as seen
   in Nov of 2009. {\em Upper Right:} FAST optical geometry. {\em Lower:} A 3D model of FAST. }
\label{fig1}
\end{center}
\end{figure}

The high frequency limit of FAST is determined by the size of the panels.
The depth of the karst depression along with the design of the suspension structure determines
the opening angle of FAST. The current surface segmentation plan and overall
constraints of budget and project time are shown to be sufficient to
realize the high frequency limit of 3 GHz and a zenith angle 40\deg.
The raw sensitivity in L-band, which is the core band for most
significant sciences of FAST, reaches 2000 m$^2$ K$^{-1}$, thanks to the huge collecting area
and up-to-date receiving system. A 19-beam feed horn receiver array  is
planned for L-band to increase the survey speed. Maximum slewing time is around 10 minutes,
which is restricted by the power of high-voltage electromotor. A summary of the technical
specifications of FAST (phase I) can be found in Table 1.

\begin{table}
  \begin{center}
  \caption{Main technical specifications of FAST.}
  \label{tab1}
 {
  \begin{tabular}{|l|}\hline
Spherical reflector: Radius = 300m, Aperture = 500m \\
Illuminated aperture: D$_i$$_l$$_l$ = 300m  \\
Focal ratio: f/D = 0.4611  \\
Sky coverage: zenith angle 40  \\
Frequency: 70MHz - 3GHz  \\
Sensitivity (L-Band): A/T $\sim$2000, system temperature T$_s$$_y$$_s$ $\sim$20K  \\
Resolution (L-Band): 2.9$'$  \\
Multi-beam (L-Band): beam number = 19  \\
Slewing time: $<$10 minutes  \\
Pointing accuracy: 8$''$  \\ \hline
  \end{tabular}
  }
 \end{center}
\end{table}

\newpage
The FAST project consists of 6 subsystems (Figure 2).
 \begin{figure}[b]
 \begin{center}
 \includegraphics[width=0.8\textwidth]{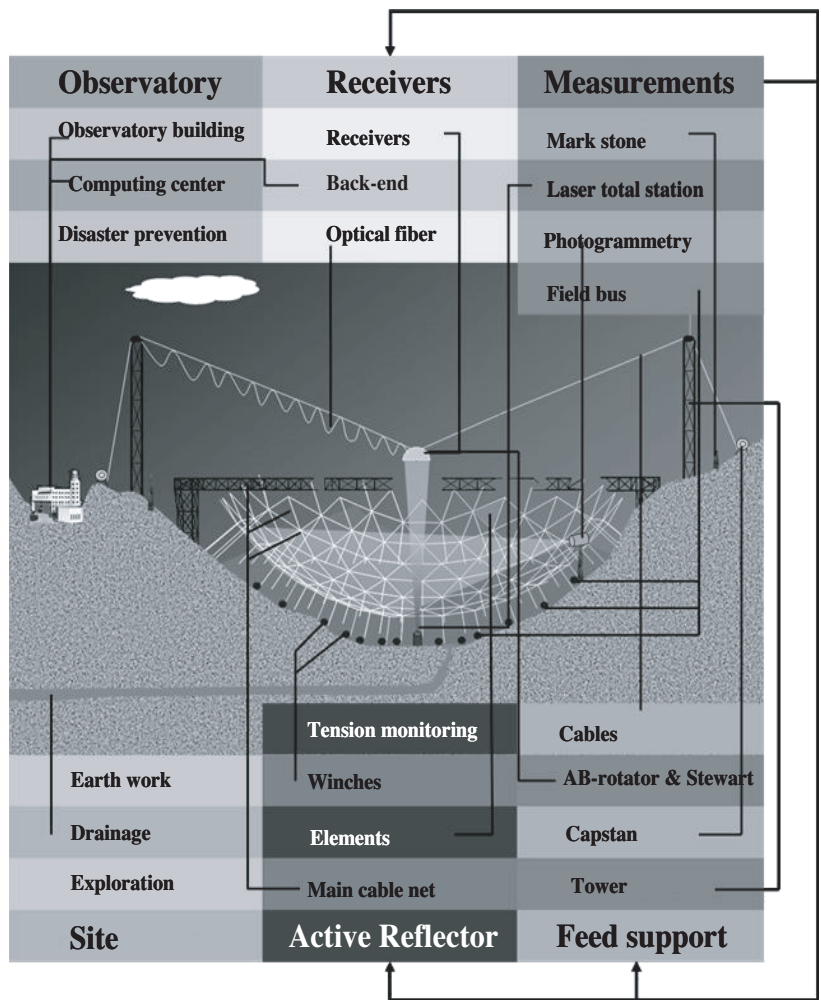}
 \caption{Illustration of the 6 subsystems of FAST.}
\label{fig2}
\end{center}
\end{figure}

\textbf{1) Site Survey and Excavation.} The natural shape of a karst formation is close
to be part of a sphere. Still, about one million cube meters of earth need to be removed,
which counts for a few percent of the total volume of a half sphere with a 500 m
diameter. As of September 2012, the site excavation and protection of the dangerous
rocks and slopes are near completion.

\textbf{2) Active Reflector.} The active main reflector, which is the most expensive part
of the project, is supported by a cable network. More than 2000 actuators drive
tie-down cables according to the feedback from the measuring system to deform the surface.
The whole reflector consists of about 4400 triangular panels, which give a RMS
error smaller than 5 mm. With a side dimension of about 11 m, the reflectors are to be
made of aluminum sheet as the surface, spatial truss of aluminum as the backup and
an adjustable layer in between. The RFI properties and the life time of actuators are
being investigated. A final selection is expected soon between mechanical and hydraulic
actuators. Cable fatigue problem poses a major challenge to the FAST design. Extensive numerical and experimental
investigation have been carried out to decrease the stress range of the cable
required. We also have developed a new type of steel strand with ultra high fatigue resistant
property. The current design and chosen materials are shown to satisfy the
operational need with sufficient reserve (Jiang 2011).

\textbf{3) Feed Cabin Suspension System.} The feed cabin of FAST is supported and
driven by cables and servomechanism without any solid structure between the cabin and
the towers. To control the position of the cabin to within the error budget, which is
the most difficult part of the FAST design, a secondary adjustable system is employed
inside the cabin. Numerous scaled models were made to verify the feasibility of the
concept. The team has also carried out end-to-end simulation through a collaboration
with MT Mechatronics and Technical University Darmstadt. These analysis show that
the displacement of feed cabin after first adjustment control can be constrained to
within a few centimeter and the secondary stabilizer further reduces the error to a few
millimeters, which meets the requirement.

\textbf{4)Measurement and Control System.} The FAST design requires a system position
accuracy of about 2 mm on a range of about 150 m, which amounts to a linear dynamic
range of 5 orders of magnitude. The measurement and control of the panels and the cabin
need to happen in real time. The 3D position and the orientation of the focus cabin will
be sampled at a rate $>$ 10 Hz. The profile of the main reflector will be measured through
1000 nodes in illuminated area in real time every couple of  minutes. The datum lines have
been established. The accuracy and stability are shown to be better than 1 mm.

\textbf{5) Receiver and Backend System.} FAST will be equipped with nine sets of receivers, covering
a frequency range of 70MHz-3GHz. Scientific backends, time/frequency standard,
and monitoring/diagnostics of the receivers are undergoing optimization.
The 19 beam feed-horn array  at L-band will be the core
survey instrument of FAST and is under development through a
trilateral collaboration among NAOC, JBCA (Jodrell Bank Centre for Astrophysics) and CSIRO (Commonwealth Scientific and Industrial Research Organisation).
The L-band single pixel receiver has been developed in the FAST lab, covering 1.1 GHz to 1.9 GHz
with return loss better than -22dB and isolation better than -22dB across the band.

\textbf{6) Observatory.} The operation center will be located in a lower depression
adjacent to the FAST site. The preliminary design has been approved. We expect the
main structure of the building to be made exclusively out of wood, which naturally fit
into the ambiance of Guizhou.

\section{Key Science Goals}
	The origin of the observable universe, the origin of our world with the Sun and the Earth, and the origin of intelligent life are overarching questions of natural sciences. FAST, with its unparalleled collecting area and its state of art receiver systems has a unique window for contribution through precise measurements of matter and energy in the low frequency radio bands.

The key science goals of FAST are based on observables between 70 MHz and 3 GHz, including
the 21 cm HI hyperfine structure line, pulsar emission, radio continuum, recombination lines, and molecular spectral lines including masers.

The majority of the normal matter in the universe is in the form of HI gas. Compared with Arecibo, FAST will have three times the scan speed at L-band and twice the sky coverage. A key goal for Galactic ISM study will be a systematic study of very cold atomic gas through measuring HI Narrow Self-Absorption features (HINSA; \cite{li03}), which will be
analyzed together with CO surveys of comparable resolution. For the local universe, FAST will conduct blind surveys to  measure the gas mass especially in optically dark galaxies. Such census of gas will help explain the discrepancy between dark matter simulation and the observable universe, in particular, the "missing satellite problem".

Using the 19 beam L-band focal plane array, FAST aims to discover over 4000 new pulsars (\cite{smits09}), about 300 of which should be milli-second pulsars. The followup timing studies of
these fast pulsars will be a substantial  addition to current pulsar timing arrays. We are looking into the
quantitive impact of FAST pulsar studies upon the detectability of gravitational wave, both from the stochastic background and single events.

FAST will  also perform targeted studies of radio continuum, recombination lines, and molecular lines.
These studies aim to enhance our understanding of the galactic structure, ISM content, and planetary physics. For example, in terms of extra-galactic maser search, FAST is expected to detect more than 1000 OH
mega-masers up to a redshift of 2 (Zhang, Li \& Wang 2011).  Such a FAST maser sample represent a
10 fold increase from the sum of all current known OH mega-masers.
FAST also stands a good chance of detecting
the most distant OH mega-masers.

 We also expect to attempt direct detection of exoplanets in meter wave band. Given the large beam of
 FAST in its lowest frequency band, the challenge of confirming the radio signal from an exoplanet will be
 systematics and stellar radio emission. We have conducted preliminary studies and plan to utilize
 the high sensitivity of FAST in that FAST will be able to sample such emission at about 10 m interval.
 The quasi-periodicity of planetary radio burst is tied to the spin of the planet, which is counted in days. The time-modulation of radio signals will much enhance the detectability of exoplanet
 by FAST.
 	
\section{Early Science Opportunities}
 We consider the first 6 months to a year after first light to be the FAST  ``early science" phase, during which a few hands-on projects will try to utilize the sensitivity of FAST before the complete suite of receivers and observing modes are successfully implemented.

The complexity and the innovative nature of the FAST systems pose many challenges. The main foreseeable one is the real-time control. With a total error budget of 2 mm, the FAST systems,
including the main active reflectors and the receiver cabin, have to be constrained precisely  through a closed feedback loop. The difficulty of realizing such precise control loop goes up substantially
with frequency.
Therefore, there exist great motivation in the early science phase to carry out projects
in wavelengths longer than the L-band. In terms of observing modes, those involving complex scan patterns and fast driving/switching should be avoided.

These two considerations discussed above mean that the early science programs should target
point sources with strong signatures below 1 GHz. Spectroscopic lines from the Orion nebular clearly
 satisfy such requirements. Our plan is to perform deep spectroscopic scan in bands as wide as possible
in low frequency ranges. There are $\lambda$ doubling line of CH, possible CH$_3$OH maser, recombination lines, and numerous other lines of more complex species in frequencies lower than
1 GHz. With a comprehensive Orion source model expected from Herschel studies, such a spectral line survey also holds the potential for discovering new lines.

We have carried out a numerical experiment to identify the optimum frequency for FAST pulsar search
in a drift scan mode. Based on an updated version of the pulsar population model by
\cite{lorimer06},  our simulations fit the detection rate of the Parkes Multibeam Survey, Parkes 70 MHz survey, and the GBT survey simultaneously by altering the spectral index and deviation of spectra index distribution together.
Our results show that the maximum detection rate for a drift-scan pulsar search by FAST can be achieved around 500 MHz, with a relatively flat 'plateau' between about 400 MHz and 700 MHz. Pulsar searches, especially toward M31, in that frequency range will thus be of high priority as a FAST early science program.

In the early science phase, FAST should  be able to detect  compact radio continuum sources, such as the
Gigahertz Peaked-Spectrum (GPS) sources. FAST will enlarge the sample to fainter flux end and shed lights into the origin of emission from these sources.

Given the technical constraints and scientific considerations regarding FAST early science operation,
two additional receivers are being studied specifically for this phase. One is a 7-beam feed horn array
 operating  between 400 MHz and 560 MHz and the other one is a single pixel receiver covering 270 MHz to 1450 MHz. The multi-beam system is being optimized for conducting pulsar searches in drift mode.
The wide-band receiver is designed primarily for line surveys. These two systems are much
cheaper and lighter compared with the suite of receivers in the formal design for normal operation.
We expect them to come online first and provide a platform for exploring the discovery space of FAST
as soon as possible. Within the technical constraints of  early science operations,
there will still be ample opportunities for
focused programs of significant impact.  Careful consideration of low frequency sources, existing surveys, and feasibility are required for   early
science programs of FAST.

The FAST group value international collaboration immensely, both in terms of
hardware design and of scientific planning and research. We expect to formulate key programs well in advance of the first light of FAST, expected in September of 2016. The detailed policies and the arrangements of the call for proposals will be formulated by an international advisory committee, which
will report to and be approved by Chinese Academy of Sciences. These key programs are
expected to be mainly lead by Chinese PIs and with substantial international collaboration, which
should be given advantages in the review process.

\section{Acknowledgement}
This work is supported by China Ministry of Science and Technology under State Key Development Program for Basic Research (2012CB821800).

\end{document}